\documentstyle[12pt]{article}
\begin{document}

\baselineskip=7mm

\begin{center}

Many-body systems in the presence
of the random interaction and the $J$ pairing interaction

\vspace{0.4in}

Akito  Arima

\vspace{0.2in}

The House of Councilors, 2-1-1 Nagatacho, 
Chiyodaku, Tokyo 100-8962, Japan

\end{center}


In this talk I shall discuss some 
regularities of many-body systems in the
presence of   random interactions and 
regularities of a single-$j$ shell for the $J$ pairing interaction
which works only when two particles are coupled to spin $J$. 
I shall first explain an empirical rule to predict the
spin $I$ ground state probability. 
 Then I shall present  
some interesting results of a single-$j$ shell under 
the $J$ pairing 
interaction.  Last I
shall discuss some  preliminary results of
binding energies in the presence of random two-body interactions. 



\newpage 
It is my great pleasure for me to talk to you here.
I would like to thank the organizers of this conference and say
``congratulations" to Professor  Stuart Pittel. 
I am also very glad to see many of my friends.

My talk includes three subjects.

  ~ (1)  spin 0 ground state dominance;

  ~ (2)  Some regularities under the $J$ pairing interaction;

  ~ (3)   ground state parities and  binding energies.

\section{spin 0 ground state dominance}

The spin $0^+$ ground state (0 g.s.) dominance was discovered and first
studied by Johnson, Bertsch, Dean, and Talmi \cite{Johnson}.
This phenomenon has attracted much  attention \cite{Zhao-review}.

In Ref. \cite{simple} we proposed a simple approach to predict
the  probability  (denoted as $P(I)$)
for a certain spin $I$ to be the ground state spin 
 without using random interactions:
 We first set one of the two-body matrix elements of the
problem to $-1$ and all the rest to zero, and then see which  angular
momentum $I$ gives the lowest eigenvalue among  all  
eigenvalues of this many-body system. If the number of independent
two-body interaction matrix elements is $N$, the above procedure
is repeated $N$ times. After all $N$ calculations 
have been done, we   count how many times (denoted as ${\cal
N}_I$) the angular momentum $I$ gives the lowest eigenvalue.
We then predict  $P(I)$  to be  ${\cal N}_I/N$. 

This approach was applied to predict the $P(I)$ 
for various systems, such as fermions in a single-$j$ shell or
many-$j$ shells, $sd$ bosons or $sdg$ bosons.  
Here we take  a single-$j$  shell with $j=9/2$ as an example. 
The Hamiltonian is defined as follows.
\begin{eqnarray}
&&  H =
 \sum_{ J} G_J A^{J \dagger} \cdot  A^{J } \equiv \sum_J \sqrt{2J+1}
\left( A^{J \dagger} \times  A^{J }  \right)^{(0)} ~,
\nonumber \\
&&  A^{J \dagger} = \frac{1}{\sqrt{2}} \left[ a_{j}^{\dagger}
\times a_{j}^{\dagger}
     \right]^{(J)} ,   ~~
     {A}^J = - (-1)^M\frac{1}{\sqrt{2}} \left[ \tilde{a}_{j} \times
     \tilde{a}_{j} \right]^{(J)}.     
\end{eqnarray}
where $G_J$ is given by $ \langle j^2 J|V|j^2 J \rangle$, and 
$V$ is a two-body interaction. Here the $J$ pairing 
interaction is defined as the interaction which has
$G_{J'} = -\delta_{JJ'}$. 

One can obtain that 
${\cal N}_0$=3 and  ${\cal N}_4$=${\cal N}_{I_{\rm max}}$=1 for
four fermions ($n=4$, $n$ labels the particle number) and 
${\cal N}_j$=2 and
${\cal N}_{3/2}$=${\cal N}_{5/2}$=${\cal N}_{I_{\rm max}}$=1 for 
$n=5$. Here  $N=j+1/2=5$. 
Thus one predicts 
that $P(0)$=60$\%$ and $P(4)=P(I_{\rm max})=20\%$ for $n=4$,
and $P(j)$=40$\%$ and $P(3/2)=P(5/2)=P(I_{\rm max})=20\%$ for $n=5$.
The $I$ g.s. probabilities obtained by using the 
random interactions are as follows: 
$P(0)$=66.4$\%$, $P(4)=11.8\%$, $P(I_{\rm max})=17.9\%$ and
other $P(I)$'s are close to zero for $n=4$;
$P(j)$=33.9$\%$ and $P(3/2)=20.5\%$, 
$P(5/2)=15.5\%$, $P(I_{\rm max})=18.4\%$ and
other $P(I)$'s are close to zero for $n=5$.
The agreement between the predicted values 
and those obtained by the random interactions is good.

There are two other conclusions based on our empirical rule. 
First,  $P(I_{\rm max})$ is considerably large for a single-$j$ shell, 
with the predicted $P(I_{\rm max})= 1/N= 1/(j+\frac{1}{2})$ which is 
 independent of  $n$. This prediction has been  
confirmed quantitatively.
Second, through the process of finding the spin $I$ of 
 the ground state of the system with $G_{J'} = -\delta_{JJ'}$,
one is able to predict  
which interactions  are essential  for a spin $I$ g.s. 
 probability.  

\section{Regularities of many-body systems under the $J$ pairing interaction}

The above empirical rule stimulated our studies of a single-$j$ shell under 
the  $J$ pairing interaction. 
To avoid confusion, here we use superscript $(n)$ to specify the particle
number $n$ for spin $I^{(n)}$  
and the eigenvalue 
$E^{(n)}_{I^{(n)}, J(j)}$ for a 
single-$j$ shell with the $J$ pairing interaction. 

We recently showed in Ref. \cite{cr8931} that
a system of three fermions in a single-$j$ shell is exactly
solvable for any $J$ pairing interaction. There is at most 
one non-zero eigenvalue for a fixed $J$
and for any $I^{(3)}$. The non-zero eigenvalue is given by
\begin{equation}
E^{(3)}_{I^{(3)}, J(j)} =  -1- 2 (2j+1) 
     \left\{ \begin{array}{ccc}
     J    & j  & I^{(3)} \\
     J    & j & j   \end{array} \right\} ~.   \label{eigen}
\end{equation}
for $G_{J'}=-\delta_{JJ'}$.

The summation of all eigenvalues over $J$ for a fixed $I^{(3)}$ is equal to
the number of spin $I^{(3)}$ states multiplied by a factor
$\frac{n(n-1)}{2}=3$. Combining this
with the result of Eq. (\ref{eigen}), one can obtain 
 a number of new sum rules for six-$j$ symbols. These sum rules
 were  found and discussed  in Ref. \cite{cr8931}.

The case of four fermions for the $J$ pairing
interaction is not soluble except for $J=0$. Here
we concentrate on the $J_{\rm max}$ pairing. In Ref. \cite{Ginocchio} we
found that many of eigenvalues $E^{(4)}_{I^{(4)}, J_{\rm max} (j)}$
are asymptotically integers (0, $-1$ and $-2$), and related them with the
number of pairs with spin $J_{\rm max}$ in their wavefunctions.
Besides these asymptotic integer eigenvalues, there are many 
 $E^{(4)}_{I^{(4)}, J_{\rm max} (j)}$ which are not close to integers.
We called them ``non-integer" eigenvalues, and noticed that they are 
very close to the non-integer eigenvalues of $n=3$. 
Therefore, it might be anticipated the wavefunction corresponding to the 
non-integer $E^{(4)}_{I^{(4)}, J_{\rm max} (j)}$ could be approximated by 
that   of the corresponding 
$E^{(3)}_{I^{(3)}, J_{\rm max} (j)}$, to which
 $E^{(4)}_{I^{(4)}, J_{\rm max} (j)}$ is  asympotically equal, 
coupled to a single-$j$ particle.
This anticipation was  confirmed by calculating their overlaps
which are 1.0 within a high precision. 
This last particle is called  a  
``spectator"  because of its weak coupling with the other strongly
bound three particles which  can be regarded as a cluster.

It is noted that the similar pattern holds for $n=5$ and 6, and also 
 for bosons with spin $l$.  The non-integer eigenvalues and
 integer eigenvalues can be 
unified by the following  picture of clusters
which include both $J_{\rm max}$ pairs and spectators:
The $J_{\rm max}$ pairing interaction favors clusters with $n_1$
particles ($n_1 \le n$), with $I^{(n_1)} \sim I^{(n_1)}_{\rm max}$.

As is well known, the existence of degeneracy indicates that the Hamiltonian
has a certain symmetry.
The degeneracy  for $J_{\rm max}$ pairing interaction, however,
is not exact. It would be interesting to explore  the broken
symmetry hidden in the $J_{\rm max}$ pairing interaction discussed
in this paper.  It would be also interesting to discuss 
the modification of the $J_{\rm max}$ pairing interaction in order to
recover the exact degeneracy.

\section{Ground state parities and binding energies} 

There are many interesting patterns of many-body systems  
in the presence of random interactions. Some characteristic
features such as the spin 0 g.s. dominance and the odd-even staggering
of binding energies  have been  numerically well known
under the random interactions.  Below I discuss two 
 other regular patterns.

The first is related to the parity distribution. In Ref. \cite{cr8930}
we studied many systems filling different  shells  
in the presence of two-body random ensemble, and 
found that  positive parity
is dominant in the ground states of  even-even nuclei, and the  
possibility  for the ground states 
of odd-mass and odd-odd nuclei  to have either positive  
or negative parity is comparable in general. This feature is very similar  
to that for the 0 g.s. dominance. Because parity is a much simpler 
quantity, the positive parity dominance might be well 
understood in the near future. This may help us to obtain 
a deeper understanding of the 0 g.s. dominance.

The second is related to binding energies obtained by the 
random interactions. The binding energy is a very important and fundamental 
quantity not only for various aspects of nuclear physics but also
for other branches of physics such as astrophysics. 
For a recent review of  nuclear  
masses, see Ref. \cite{bind-review}. Recently Bohigas and Leboeuf
claimed in Ref. \cite{chaos} that 
there is a lower limit on the accuracy of theoretical predictions of binding
energies due to chaotic components in the nuclear mass.
Here I present some preliminary results for binding energies of 
a few neighboring nuclides  in the $sd$ shell  which are  
obtained by using the random interactions. 
 We are interested in 
the Garvey-Kelson relations  which connect
the masses of neighboring nuclides.  Our calculated results seem to
satisfy the algebraic  Garvey-Kelson relations   within a
reasonable precision in the presence of random interactions.  
For instance, we calculated the deviations from the Garvey-Kelson 
relation for example around $^{22}$Na, which involves of 
  binding energies of Na$^{21,22}$,
Mg$^{21,23}$ and Al$^{22,23}$, and found that  
these deviations are   small.
More work is necessary to clarify the indications
of Ref. \cite{chaos}.

\section{Summary}

In this talk I have discussed a few interesting
aspects concerning the regularities of many-body systems
under the random interactions and a 
single-$j$ shell under the $J$ pairing interaction. 
First, I explained our empirical rule which is useful in  predicting
the $I$ g.s. probabilities   and understanding   the spin  
0 g.s. dominance. Second, I showed that  three fermions in a
single-$j$ shell for any $J$ pairing interaction is solvable. 
I also discussed  more fermions among which the 
$J_{\rm max}$ pairing interaction works, and suggested  
that the  $J_{\rm max}$ pairing interaction favors clusters, where 
$I^{(n_1)}$ ($n_1 \le n$) for each cluster is 
very close to or equal to 
the maximum spin for $n_1$ particles. 
Last,  I showed two other interesting
features for the ground state parities and
binding energies by using the random interactions:  ~(1) 
the positive parity dominates in the ground
states of even-even nuclei, and   (2) the calculated binding energies satisfy 
 the Garvey-Kelson relations
reasonably well.

Acknowledgement: I would like to thank Dr.
Y. M. Zhao for his  collaboration in this work.
I also thank Dr. I. Talmi for his valuable comments
on some results of Sec. 2.

\end{document}